# Microscopic approaches for understanding the mechanical behaviour of reinforcement in composites

Dr D. Durville, MSSMat Laboratory, Ecole Centrale Paris / CNRS UMR8579
Grande Voie des Vignes, 92290 Châtenay-Malabry, France
Email: damien.durville@ecp.fr

**Abstract**: an approach to the mechanical behaviour of textile composites at the scale of their constituting fibres, using an implicit finite element simulation code, is proposed in this chapter. Based on efficient methods and algorithms to detect and take into account contact-friction interactions between elementary fibres, it allows to model samples of textile composites made of few hundreds of fibres. The approach is employed first to determine the initial configuration of woven fabric samples, before applying to them different loading cases in order to identify their mechanical properties under various sollicitations. A modeling is proposed to join an elastic matrix to the woven structure by the means of nonconforming meshes and connection elements, in order to simulate the application of loading cases to samples of woven composites. Addressing the problem at the microscopic scale of fibres the approach does not require the identification of models at intermediate scales.

## 15.1 Introduction

Textile composites manufactured from woven fabrics are formed of arrangements of fibres and matrix. The organization of fibres into two levels – the tows and the woven fabric – gives a a multiscale aspect to these materials, characterized by a complex mechanical behaviour whose nonlinearities are mainly due to the interactions between components. Various modelings of this mechanical behaviour are possible depending on the scale at which phenomena are addressed. Macro-models at the scale of the fabric require an homogeneized approach to the fabric behaviour. At a lower scale, having identified meso-models representing the behaviour of elementary tows, it is possible to model the woven structure as an assembly of tows, taking into account interactions between them. Models at intermediate scales can not yet account for phenomena occuring at the scale of fibres, such as local compaction or rearrangement of fibres within tows, and which can be responsible for nonlinear effects at the global scale.

The approach presented here aims at modeling samples of textile composites at the scale of fibres, using a simulation code based on the finite element method, which was specially developed for the study of the mechanical behaviour of entangled materials, in order to identify their mechanical response when subject to various types of loadings (Durville 2005, 2009, 2010). The key issue in this approach at microscopic scale, which considers all fibres constituting the studied samples, is the taking into account of contact-friction interactions developed between these fibres. The solving of the mechanical problem involving a large number of contacts in a reasonable computation time, and using an implicit solver, is made possible by the development of methods and algorithms to detect contacts and to handle the nonlinear interactions they generate.

The development of such methods allows to handle assemblies made of few hundreds of fibres, and subject to large displacements and strains, and to simulate the behaviour of small samples of woven fabrics directly at the scale of fibres. Two main results are derived from this approach. The first one is the computation of the initial configuration, using a special process which gradually moves fibres until satisfying the chosen weaving pattern. The trajectory of each fibre is thus obtained through the solution of a global mechanical equilibrium, which offers an accurate geometrical description of the arrangement of fibres and tows within the fabric. In a second time, this woven structure can be submitted to different loading cases, to identify its mechanical behaviour.

This way of tackling the problem at the scale of fibres has two main advantages. First of all it eliminates the need for the definition of the initial configuration, which can be very complex. Secondly, no intermediate model at meso-scale is required, and the only parameters to be defined are the characteristics of fibres (radius and elastic parameters) and the parameters governing contact-friction reactions.

As the fabric is joined to an elastic matrix to form the composite, a particular modeling is made to account for the presence of this matrix and for its interactions with fibres. A solid mesh, non-conforming with the meshes of fibres, but with an overlapping region of controled thickness with the tows, is automatically generated on both sides of the fabric. Connection elements are then introduced in the overlapping region to ensure the link between the fibres and the matrix. This modeling allows to simulate the application of loading cases inducing large displacements on composite samples, coupling the approach at the mesoscopic scale of fibres with the consideration of an elastic matrix.

To present the proposed approach, the chapter is organized as follows. The next section describes the interests of the approach at microscopic scale.

## 15.2 Interests and goals of the approach at microscopic scale

### 15.2.1 Complexity of textile composites reinforcements

Because they are made of arrangements of fibres, tows and matrix, textile composite reinforcements appear as heterogeneous materials with a multi-scale structure. If they can be considered as continuous structures at a macroscopic scale, at lower levels, they must viewed as discrete assemblies of tows and fibres. Their global behaviour thus involves both continuous aspects, essentially related to the behaviour of fibres and tows in longitudinal directions, and discrete aspects related to interactions between tows or fibres, and influencing mostly the behaviour of these components in transverse directions.

Due to their low transverse and bending stiffnesses, these materials may undergo large deformations which can change the geometry and the arrangement of tows and fibres, thus inducing couplings and nonlinearities at different levels.

## 15.2.2 Determination of the initial geometry

An accurate definition of the geometry of woven fabrics used as reinforcement may be essential for two reasons. First, a good knowledge of porosities between tows and fibres is required to assess the permeability of such materials with respect to the injection of resin during the impregnation process. Second, running finite element simulations at the meso scale in order to assess the mechanical response of textile composites requires to be able to describe the geometry of tows in order to mesh the different components corresponding to the tows and the matrix.

However, the arrangement of fibres and tows in a woven assembly is hard to determine a priori since it results from complex interactions between these components during the weaving process. The way fibres rearrange within tows, and the way tows deform accordingly in different directions, both play a predominant role in the determination of the trajectories of fibres and tows, and of the shapes of tows cross-sections.

Geometrical approaches have been developed to model the initial geometry of textile materials. Robitaille and al. (2003) developped an approach to determine volumes occupied by matrix and tows in an unit-cell, by identifying elementary volumes based on geometrical constructions. Verpoest and Lomov (2005) approximate the trajectories of yarns by geometries minimizing the yarn deformation energy, and generate the volume of yarns by sweeping these trajectories with crosss-sections of constant shape but with varying sizes. Hivet and Boisse (2008) try to identify precisely the varying contours of cross-sections along the yarn, depending on their relative positions with respect to the other yarns, to provide an accurate geometrical description of the woven structure, avoiding in particular penetration between tows. All these approaches are designed to supply mechanical models, and particularly finite element models, with a relevant geometrical description of components of a textile composite. Geometrical parameters on which these approaches are based require to be identified from observations on manufactured fabrics.

An alternative way to approach the geometry of textile reinforcements is to simulate the manufacturing of these structures so that their components (yarns or fibres) take naturally their equilibrium positions depending on the interactions they develop with other components. By this way, only very few geometrical assumptions are needed. Dynamic explicit FE simulation codes have been used in particular for this purpose. Finckh (2004) simulated the weaving of fabrics, taking into account several fibres within each yarn. Pickett *et al.* (2010) simulated the braiding of a preform made of 96 yarns, representing each yarn by bar elements, using also an explicit FE simulation code. Simulation by means of so-called digital elements was proposed by Miao *et al.* (2008) and Wang *et al.* (2010) to calculate the geometry of woven and braided structures, considering several fibres within each yarn.

Similarly, the approach we propose here aims at determining the initial geometry of samples of woven fabric, by making tows made of several fibres fulfil gradually the weaving pattern chosen for the fabric. This approach, characterized by an implicit method to solve the mechanical equilibrium, is based on methods and tools specially developed for the mechanical modeling of entangled materials.

## 15.2.3 Approach to the mechanical behaviour of textile reinforcements

Beside geometrical issues, textile composites are characterized by complex mechanical behaviours at different scales. Tows display a very specific mechanical behaviour due to the fact they are formed of an assembly of discrete fibres. Their behaviour in longitudinal direction mainly depends on the elasticity of fibres, whereas their behaviour in transverse directions is essentially ruled by contact-friction interactions between fibres. Friction furthermore introduces couplings between loadings in different directions since frictions forces between fibres depend on normal reactions, which depend themselves both on the curvature of fibres and on their tensile stress. Because of these friction effects, the twist of tows and their tensile stress have a large influence on their transverse behaviour.

The scale of tows can be qualified as meso scale. Identifying non linear and coupled behaviours at this meso scale is a difficult task which requires a sound understanding of phenomena taking place between fibres. Although many attempts have been made to propose such meso models (see for example Lomov *et al.*, 2007, Boisse *et al.*, 2010), this remains an open issue.

Taking into account the individual fibres constituting the tows, the formulation of models at the meso scale is no longer necessary. Only the mechanical behaviour of fibres, the geometrical arrangement of these fibres and the friction interactions between them need to be defined to represent the global behaviour of the tow.

## 15.3. Modeling approach to textile composites at microscopic scale

To identify the mechanical behaviour of textile composites, small samples made of fibres and matrix, and subject to various loadings on their edges are studied. The global problem is set in the form of the seek of the mechanical equilibrium of an assembly of fibres, under quasistatic assumptions, and considering finite strains and large displacements.

To achieve such a modeling, the mechanical behaviour of each constituent component, namely the fibres and the matrix, have first to be accounted for through appropriate models. In a second time, interactions between these components (contact-friction interactions between fibres and connections between the fibres and the matrix) need to be represented. Finally, as the edges of the considered samples consist of assemblies of fibres, a complex driving of boundary conditions must be implemented in order to apply the desired loadings without interfering too much with the changes of the local arrangement of fibres within tows.

### 15.3.1 Finite strain beam model for fibres

An enriched kinematical beam model has been adopted to represent the behaviour of fibres. According to Antman's theory (Antman 2004), this model describes the kinematics of any beam cross-section by the means of three vector fields, one for the position of the centre of the cross-section, and two directors to represent first order deformations of the cross-section. According to this model, the position of any particle $\boldsymbol{\xi}$ of the beam, identified by its three components $(\xi_1,\xi_2,\xi_3)$ in a material configuration, can be expressed as follows :

$$\mathbf{x}(\xi_1,\xi_2,\xi_3) = \mathbf{x}_0(\xi_3) + \xi_1 \mathbf{g}_1(\xi_3) + \xi_2 \mathbf{g}_2(\xi_3), \qquad [1]$$

where $\mathbf{x}_0(\xi_3)$ is the position of the center of the cross-section, and $\mathbf{g}_1(\xi_3)$ and $\mathbf{g}_2(\xi_3)$ are two directors of the section (Fig. 15.1), these three vectors depending only on the curvilinear abscissa $\xi_3$. This expression can be interpreted as a first order Taylor expansion of the position vector with respect to the transverse coordinates $(\xi_1,\xi_2)$ of the particle. Accordingly to the expression of the position, the displacement of any particle is written in the following way :

$$\mathbf{u}(\xi_1,\xi_2,\xi_3) = \mathbf{u}_0(\xi_3) + \xi_1 \mathbf{h}_1(\xi_3) + \xi_2 \mathbf{h}_2(\xi_3), \qquad [2]$$

where $\mathbf{u}_0(\xi_3)$ is the displacement of the center of the cross-section, and $\mathbf{h}_1(\xi_3)$ and $\mathbf{h}_2(\xi_3)$ are the variations of the section directors.

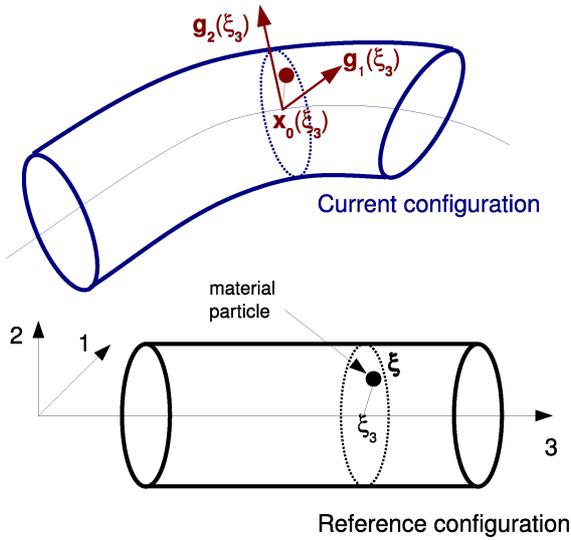

Figure 15.1 : beam kinematical model

No assumption is made on the norm of the section directors, nor on the angle between them. This means that the cross-sections of the beam can deform depending on the variations of the section directors. However, as the surfaces of cross-sections are generated from the two directors, they remain plane when deformed. If their initial shape is circular, they can deform to any elliptical shape.

This kinematical model considers nine degrees of freedom per cross-section, and is therefore a little more expensive than models assuming rigid cross-sections, and using only six degrees of freedom per cross-section. This enriched kinematical model however offers some advantages. First, the use of directors to define the kinematics of cross-sections avoids the handling of finite rotations and greatly simplifies the formulation of the model. Second, thanks to the deformations of cross-sections introduced by these directors, no component of the Green-Lagrange strain tensor is a priori zero. This allows to use standard constitutive laws involving all components of the strain tensor, and to reproduce in particular the Poisson's effect.

**Adaption of stiffness to model macro-fibres**

The actual number of fibres in each tow may be too high for the problem to be solved in reasonable time. To circumvent this limit, a reduced number of so-called macro-fibres, with a larger diameter, can be considered. If $N_f$ is the actual number of fibres in a tow, with diameter $D_f$, and $N_m$ is the number of macro-fibres chosen to represent these actual fibres, the diameter $D_m$ of macro-fibres is calculated to keep the same cross-sectional area :

$D_m = (N_f / N_m)^{1/2} D_f$.  [3]

Since the axial stiffness of a fibre is proportional to its cross-sectional area, a macro-fibre will have an axial stiffness equivalent to the one of the number of actual fibres it represents. However, as the bending and torsional stiffnesses depend on the second moment of area, $I_f = \pi r^4/4$, which is proportional to the radius to the power 4, the bending and torsional stiffnesses of the macro-fibres would be overestimated. To correct this effect, the second moment of area of the macro-fibre, denoted $I_m$, is calculated as

$$I_m = N_f / N_m \, I_f. \qquad [4]$$

## 15.3.2 Contact-friction interactions between beams : a central issue

The taking into account of contact-friction interactions is the main issue in the problem considered here. The high number of fibres leads to a high density of contacts between fibres, and the modeling of contact and friction interactions introduces additional nonlinearities.

The consideration of these nonlinearities is what makes the approach relevant to follow phenomena at the scale of fibres, but complexifies the solving of the problem. Effective methods are required to detect contact, and appropriate models for contact and friction, leading to efficient numerical algorithms, must be developed to reduce the number of iterations needed per loading step using an implicit solver.

**Geometrical aspects : detection of contacts within an assembly of fibres**

Due to the large deformations undergone by the tows, contacts between fibres change continuously during the loading. Not only their relative locations with respect to the fibres can change, but some of them can appear and disappear. As contacts are not fixed, their detection must be regularly updated, and the method employed to detect needs to be effective.

The goal of the detection of contact is to associate entities on the surface of fibres to which non interpenetration conditions should be applied. Because of the one-dimensional geometry of fibres, we consider that a contact zone between two fibres is represented by a line. Depending on the relative orientations between the two fibres, the contact line formed between them can have very different extensions : for two perpendicular fibres, this contact line reduces to a point, whereas this line can have the same length as the fibres when they follow parallel trajectories. In order to account for these different configurations, and consistently with the approximation of the geometry of fibres by finite element shape functions, we propose to check contact between fibres by the means of punctual contact elements constituted by pairs of material particles that are predicted to enter into contact.

*Generation process of contact elements*

The generation of contact elements often follows a master-slave type strategy. Determining one point on the surface of the slave structure, a corresponding target is searched on the opposite master surface, using a given contact search direction, usually taken as the normal vector to one of both surfaces. This method can be criticized for not providing a symmetrical treatment for both structures, due to the fact that the contact search direction is determined from the geometry of only one of both structures.

Instead of determining contact from one structure with respect to the other, the approach taken here consists in considering both structures with respect to an intermediate geometry defined as an approximation of the actual contact line. By this way, the contact search direction is determined from this intermediate geometry, and both structures are considered symmetrically.

The method is implemented according to the following steps. First, proximity zones, defined as pairs of parts of fibres that are close enough to each other, are determined within the whole assembly of fibres. In order not to be too time consuming, this search of proximity zones is performed evaluating distance criteria at test points distributed on the fibres with a coarse discretization. The k-th proximity zone between fibres i and j, denoted $Z^{ij}_k$, is defined as follows by a pair of intervals of curvilinear abscissae on both fibres (see Fig. 15.2) :

$$Z^{ij}_k = \{[a^i,b^i],[a^j,b^j]\}. \quad [5]$$

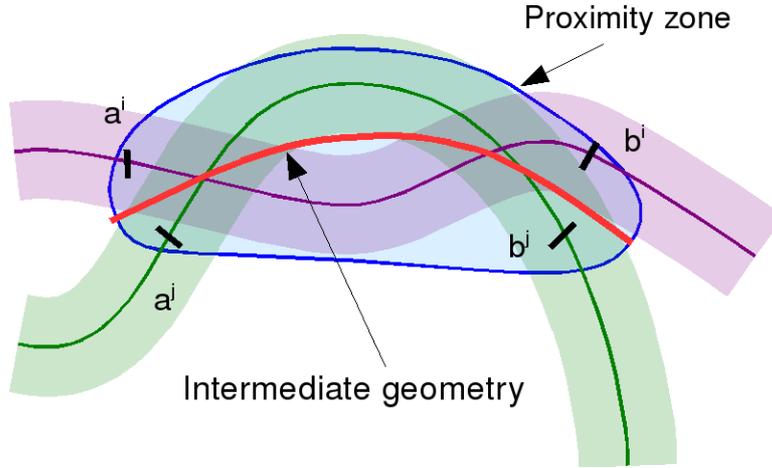

Figure 15.2 : determination of a proximity zone between two fibres

For each proximity zone, the intermediate geometry is defined as the average of both parts of line constituting the proximity zone. Each point on this geometry, identified by its relative curvilinear abscissa s, is defined by

$$\mathbf{x}_{int}(s) = [\mathbf{x}_0^i(a^i + s(b^i - a^i)) + \mathbf{x}_0^j(a^j + s(b^j - a^j))] / 2. \quad [6]$$

This average geometry, viewed as a means to approximate the unknown actual geometry of the contact line, is then employed as geometrical support for the discretization of contact. Contact will be checked at some discrete points xc defined by their abscissa sc on the intermediate geometry :

$$\mathbf{x}_c = \mathbf{x}_{int}(s_c). \quad [7]$$

For each discrete point on the intermediate geometry, a contact element is defined as the pair of material particles located on the surface of fibres that is predicted to enter into contact at this location :

$$E_c(\mathbf{x}_c) = (\xi^i, \xi^j) \text{ such } \xi^i, \xi^j \text{ enter into contact at } \mathbf{x}_c. \quad [8]$$

The determination of the two material particles candidate to contact is performed in two steps (see Fig. 15.3). First we determine the position of the centers of cross-sections candidate to contact by the intersections between the plane orthogonal to the intermediate geometry going through $\mathbf{x}_c$ and the centerlines of the fibres. Next, for each cross-section, the particle candidate to contact is localized on the outer contour of the section using the directions between both centres of cross-sections.

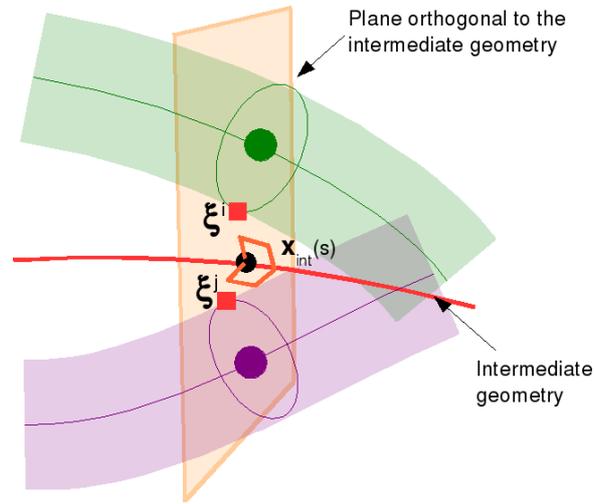

Figure 15.3 : determination of material particles constituting a contact element

*Linearized contact conditions*

Once pairs of material particles constituting contact elements have been determined, kinematical conditions designed to prevent interpenetration between fibres must be established for these contact elements. A gap distance gap ($E_c$) is defined between these particles according to a normal contact direction $\mathbf{N}(E_c)$ :

gap ($E_c$) = ($\mathbf{x}^i(\xi^i) - \mathbf{x}^j(\xi^j), \mathbf{N}(E_c)$). [9]

In order to fulfill non interpenetration conditions, the gap distance defined for each contact element is prescribed to be positive. The determination of the normal direction $\mathbf{N}(E_c)$ is essential in the formulation of this condition. This direction needs to be properly oriented in order to prevent interacting fibres from crossing through each other. Depending on the angle between fibres, it is calculated in different ways using the positions of the centers of cross-sections and the tangent vectors to the centroidal lines.

**Mechanical models for contact and friction**

The convergence of algorithms used to solve contact problems depends to a large extent to the mechanical models expressing contact-friction interactions at contact elements.

*Mechanical for normal contact*

To enforce kinematical contact conditions, a penalty model is employed, but with two main refinements. The first one concerns of a regularization of the standard penalty method, by a quadratic function for very small penetrations. Instead of defining the normal reaction as simply proportional to the gap, with a penalty parameter $k_N$,, it is assumed to be quadratic with respect to the penetration for penetrations lower than a given regularization threshold $p_{reg}$, and linear for larger ones, according to the following expression :

if gap ($E_c$) ≥ 0,    $R_N$ ($E_c$) = 0 ,

if $-p_{reg}$ ≤ gap ($E_c$) ≤ 0,    $R_N$ ($E_c$) = ($k_N/(2\ p_{reg})$) (gap ($E_c$))$^2$,

if gap ($E_c$) ≤ $p_{reg}$,  $R_N$ ($E_c$) = - $k_N$ gap ($E_c$) - ($k_N$ / 2) $p_{reg}$. [10]

The regularization ensures a continuity of the derivative of the normal reaction with respect to the gap at the origin. This improvement is very useful to stabilize the contact algorithm, particularly for contacts with very low reaction force, the status of which may be oscillating from an iteration to the next.

The second improvement of the penalty method is provided by a local adaptation of the penalty parameter for each proximity zone. The determination of this parameter is a delicate issue since it controls the penetration allowed by the penalty method at contact elements. The quadratic regularization is efficient provided a small amount of contact elements is concerned by the regularization, and thus have penetrations under the regularization threshold. However, since contact loads may vary widely from one contact zone to the other, and during the loading, a constant and uniform penalty parameter can not ensure penetrations of the same order of magnitude in different contact zones, as required to guarantee the efficiency of the quadratic regularization. This is the reason why the penalty parameter is adapted for each proximity zone, so that to control the maximum allowed penetration within each contact zone. With this adaption, penetrations registered at different contact elements within a given proximity zone are different, depending on the load exerted at each contact element, but the maximum penetration is approximately limited by the maximum allowed penetration.

*Mechanical model for friction*

A regularized friction model, based on the Coulomb's law, and allowing a reversible relative displacement before gross sliding occurs, is used to account for tangential interactions between fibres. The fact that contact elements have no continuity in time requires an appropriate procedure to transfer history variables related to the reversible relative displacement from one step of computation to the following, between different contact elements. For this purpose informations related to this quantity for each particle of the contact element are attached to material configuration, and are interpolated at the location of the contact element each time contact elements are updated.

**Algorithmic aspects**

The modeling of contact-friction introduces various nonlinearities in the global problems that need to be solved by appropriate algorithms. The first nonlinearity is related to the generation of the contact elements which clearly depends on the solution. As the solution can change significantly during one loading increment, the positions of contact elements should updated. The relations defining the positions of particle of contact elements can not be differentiated, which prevents the use of a Newton-like algorithm. A fixed point algorithm is therefore dedicated, at a first level, to iterations on the determination of contact elements, within each loading increment. Contact elements being fixed, another fixed point algorithm is used to iterate on the normal contact directions involved in the expression of the gap function. All other nonlinearities (contact status, sliding, finite strains) are treated by a Newton-Raphson algorithm. The nesting of loops of three levels tends to increase the number of total iterations, but is necessary to consider large loading increments. Efficient algorithms are all the more needed to solve nonlinearities related to mechanical models for contact and friction.

### 15.3.3 Modeling of the matrix and its interactions with textile reinforcements

Composite parts are of made of a combination of textile reinforcements and matrix, and both components must be considered to model the behaviour of a textile composite sample. We consider here composite samples made of the juxtaposition of two matrix layers on both sides of the fabric. The exact determination of the volume occupied by the matrix is a difficult issue since it depends on the manufacturing process employed and on the way the matrix impregnates the tows. An accurate geometrical description of this matrix volume may contain many details of very small scale, and meshing such a geometry can turn out to be a difficult task, yielding a large number of small finite elements that could significantly increase the computational cost.

To overcome this difficulty, we choose to approximate the actual geometry of the matrix in order not to take into account small details. The matrix is assumed to penetrate textile tows with a given penetration depth, and a solid structured mesh is generated to approximate the geometry of both layers. This meshing is performed independantly on the finite element discretization of fibres. The resulting meshes for the matrix and the fibres are thus non conforming. The continuity of displacements on the interface between these components is therefore no longer ensured by the sharing of common nodes, and an additional developments are required to model the binding between these structures. Special bonding elements are introduced for this purpose.

This global modeling of the matrix in interaction with the fabric tries to reproduce first order effects of the coupling between textile reinforcements and the matrix, focusing on phenomena occuring within the textile components of the composite.

**Meshing of the matrix**

The meshing of the matrix is performed once the initial configuration of the woven fabric has been computed, providing an accurate description of the volume occupied by fibres.

In order to mesh the volume of each layer, a regular mesh is first carried out on the external face of the layer. Each node of this face is then projected vertically on the first fibre encountered on the fabric. This projection point is then moved to penetrate with a given depth inside the fabric. A structured solid mesh for the layer is finally generated between these two faces (see Fig. 15.4).

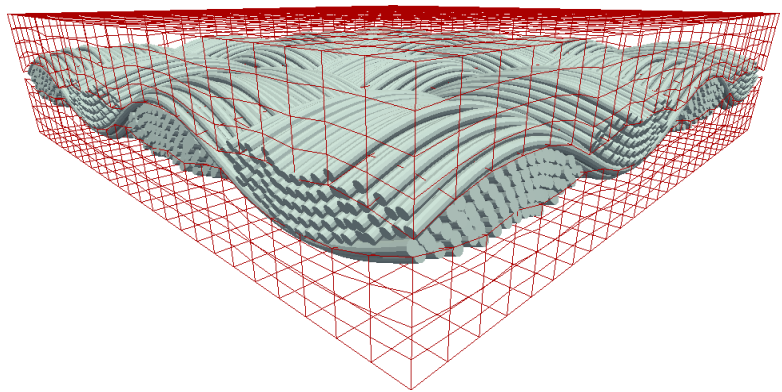

Figure 15.4 : mesh of the volume occupied by the matrix generated on both sides of the fabric

This way of meshing the matrix is pretty simple and does not require any particular geometrical tool. The penetration of the internal surface of the matrix layer inside the fabric provides an overlapping region between the two components within which the binding between both structures will be modelled.

**Introduction of bonding elements between the fibres and the matrix**

Because the nonconforming meshes for the fibres and the matrix do no longer ensures the continuity of displacements at the interface, a particular modeling of the coupling between these two structures is required. Particular bonding elements are introduced for this purpose at nodes of fibres which are located inside solid elements of the matrix. For each fibre node $I_f$, located inside a finite element $E_m$ of the matrix, we seek the corresponding material particle $\xi$, defined by its relative coordinates $(\xi_1, \xi_2, \xi_3)$ within the matrix finite element, whose position $\mathbf{x}(\xi)$ is characterized by:

$$\mathbf{x}(\xi) = \sum_{J,m} w_{J,m}(\xi_1,\xi_2,\xi_3) \, \mathbf{X}_{J,m} = \mathbf{X}_{I,f}, \tag{10}$$

where $(J_m)$ are the nodes of the solid finite element of the matrix, $w_{J,m}$ are the shape functions associated to these nodes, and where $\mathbf{X}_{I,f}$ and $\mathbf{X}_{J,m}$ are respectively the position of the considered fibre node, and the positions of the nodes of the solid element of the matrix.

For each fibre node located inside a matrix finite element, a bonding element is generated between this node and the corresponding material particle in the matrix volume, by introducing a bonding stiffness $k_b$ between the two points. The resulting interaction force between the two points writes

$$\mathbf{R}_b = k_b \left( \sum_{J,m} w_{J,m}(\xi_1,\xi_2,\xi_3) \, \mathbf{X}_{J,m} - \mathbf{X}_{I,f} \right). \tag{11}$$

The introduction of the virtual work of this force induces couplings between degrees of fredom associated to the fibre node and to the matrix nodes of the corresponding matrix element.

The bonding stiffness $k_b$ must be chosen carefully. As the displacement fields in the matrix finite elements are interpolated with a much coarser scale than they are at the level of fibres, and as several fibres can be bound within the same matrix finite element, taking a too strong bonding stiffness would amount to prescribe a coarse kinematics to the set of fibres involved in the matrix finite element, and to lock somehow to possible relative motions between these fibres. In order to give back more flexibility to the connection, by allowing it some extension, the bonding stiffness is computed for each bonding element in function of the Young's modulus of the matrix and of the size of the finite elements of the fibre and the matrix.

## 15.3.4 Modeling of boundary conditions

**Problems with boundary conditions**

Dealing with boundary conditions to apply at the ends of the tows and of the fibres on the edges of the studied sample, we are confronted with a contradiction. On the one hand we would like to apply displacements at the ends of the fibrous components on the edges of the sample to maintain it, whereas on the other hand, fibre ends and tow ends should be allowed to rearrange according to the loading to which they are subjected.

**Hierarchical organization of the composite**

The hierarchical organization of the composite sample in different levels related to the fibres, the tows, the woven fabric and the matrix, must be respected by the application of boundary conditions, to allow to apply simultaneously different conditions to the ends of components from different levels. For example, one may wish to fix the end of a tow while allowing the ends of its constituting fibres to move.

**Introduction of rigid bodies at ends of components**

To apply different conditions according to the different hierarchical levels of components, we introduce rigid bodies at their ends. Boundary conditions can then be applied to these rigid bodies. Ends of subcomponents can then be linked with proper conditions to the rigid body associated to the upper level. Rigid bodies are generated for each edge of the sample, and for each end of the tows. By this way, ends of tows on one edge of the sample can be attached with appropriate connection conditions to the rigid body associated to this edge, and ends if fibres constituting a tow can be attached to the rigid body associated to its end.

The connection conditions established between a set of ends and a rigid body must enable to prescribe globally a given displacement or a given force to this set of ends, while allowing them the possibility to rearrange. Average connection conditions have been developped to meet this need.

The employed rigid bodies are defined by four nodes ($N_m, N_{D1}, N_{D2}, N_{D3}$) (see Fig. 15.5), the first node being the master node of the rigid body, and the other three forming an orthonormal frame. These four nodes are rigidly connected to maintain the orthonormality of the frame expressed by the following conditions :

$$(\mathbf{x}(N_{Di}) - \mathbf{x}(M), \mathbf{x}(N_{DJ}) - \mathbf{x}(M)) = \delta_{ij}. \qquad [12]$$

To prescribe in average a displacement $d_i$ in the i-th direction of a moving rigid body to a set of $n_e$ ends, represented by the nodes ($N^e_k$), we express the following condition :

$$1/n_e \sum_k \left(\mathbf{x}(N^e_k) - \mathbf{x}(M), \mathbf{x}(N_{Di}) - \mathbf{x}(M)\right) = d_i. \qquad [13]$$

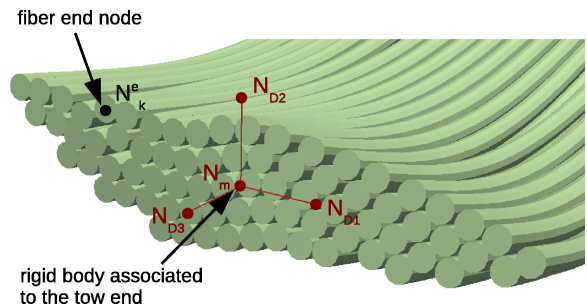

Figure 15.5 : rigid body use to define boundary conditions at the end of a tow.

Since the rigid body to which end nodes are attached is moving, this average conditions is nonlinear. Nodes involved in such a conditions are constrained only in average, and each particular node can move differently, provided the mean displacement corresponds to the prescribed value.

The use of the average connection conditions together with the introduction of moving rigid bodies at the ends of tows and on the edges of the sample is essential apply appropriate boundary conditions on the edges of the sample.

## 15.3.5 Determination of the initial configuration of the woven fabric

The knowledge of the initial configuration is a key issue for woven fabric. As the arrangement of fibres and tows within the fabric results from the different stages of the manufacturing process, trajectories of fibres and tows making up the fabric can hardly be known a priori. One way to access by calculation to these geometries could be to simulate the whole manufacturing process, and in particular the weaving process. Such a simulation would require to consider significant lengths of yarns, which would be too expensive using an implicit solver. Simulations of weaving and braiding processes have been proposed (Finckh 2004 Pickett et al. 2009) using explicit dynamic simulation codes.

To avoid a too expensive simulation of the weaving process, we propose an alternative method consisting in moving progressivelly tows forming the fabric until their arrangement fulfill the chosen weaving pattern. To do this, we start from an arbitrary initial configuration, where all tows, constituted by a compact arrangement of fibres, are in the same plane, interpenetrating each other (see Fig. 15.6). From the weaving pattern, at each crossing between tow tows, it is possible to state which tow should be above or below the other.

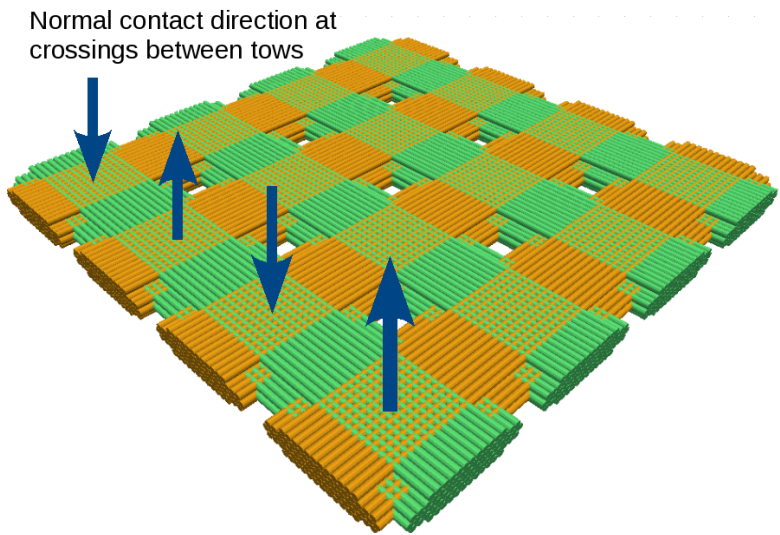

Figure 15.6 : starting configuration for the determination of the initial geometry of the woven fabric, and normal contact directions transiently chosen for fibres at crossings between warp and weft tows for a plain weave

During the stage of determination of the initial configuration, contact conditions between fibres are modified so that fibres of initially intersecting tows are gradually moved in the right direction until there is no more penetration between tows. Two adjustements to the contact conditions are made for this purpose. The first one concerns the choice of the normal contact condition. As we know, from the weaving pattern, which fibre should be above the other at each crossing between tows, this normal contact direction is oriented to statisfy this condition. The second adjustement is related to the evaluation of the gap between particles of contact elements. The observed penetrations during this initial stage have no physical meaning and can be very important. As the treatment of contact conditions tends to reduce the penetrations to zero, this would lead to too important relative motions between fibres, preventing the convergence of the global solution algorithm. To make the fibres move gradually, the gap considered between two particles is limited to a given value. The relative displacement between the two concerned fibres thus does not exceed this value. The global increment of displacement for each loading step during this initial stage can then be controlled to guarantee the convergence of the solution algorithm.

## 15.4 Application examples

Different results obtained on the same intial set of tows, arranged according two different weaving patterns, and subject to various loading cases, are presented in this section. Main characteristics of this initial set of tows, geometrical characteristics of the fabric and mechanical characteristics of fibres are respectively summarized in Tables 15.1, 15.2 and 15.3.

Table 15.1 : characteristics of the model

| | |
|---|---:|
| Number of tows in weft direction | 4 |
| Number of tows in warp direction | 4 |
| Number of fibers per tow | 76 |
| Total number of fibres | 608 |
| Number of finite elements per fibre | 32 |
| Total number of nodes | 39 520 |
| Total number of degrees of freedom | 355 680 |
| Estimated number of contact elements | 100 000 |

Table 15.2 : geometrical characteristics of the fabric

| | |
|---|---:|
| Radius of fibres | 0.0201 mm |
| Tows cross sectional area | 0.0967 mm$^2$ |
| Initial distance between tows (before crimp) | 0.88 mm |

Table 15.3 : mechanical characteristics of fibres

| | |
|---|---:|
| Young modulus of fibres | 7300 MPa |
| Poisson's ratio | 0.3 |
| Adjustement of the second moment of area for fibres | 0.20 |
| Friction coefficient | 0.15 |

### 15.4.1 Determination of the initial configuration of the woven structures

The first task assigned to the proposed approach is the determination of the a priori unknown initial configuration of the woven structures. Starting from the same initial arrangement of eight tows, each made of 76 fibres, two different patterns of plain weave and twill weave are chosen to compute this initial geometry. For this first stage of simulation, 21 increments are needed. Tension forces are applied in both directions, before being reduced to very values during the last increments. The obtained initial configurations are shown in Fig. 15.7 and 15.8. Useful geometrical informations can be derived from these results, in particular the crimp of the fabric, and the local curvatures in all fibres. The determination of the trajectory of each individual fibre within the fabric allows to describe the trajectories and cross-sections shapes of tows. To illustrate the method of determining the initial configuration, Fig. 15.9 shows the evolution of the shapes of the tows during this firts stage. Fig. 15.10 compares slices of the computed initial configuration for the two different weaves and shows different cross-sections shapes. These descriptions can be of great interest for the construction of meso-models at the scale of tows.

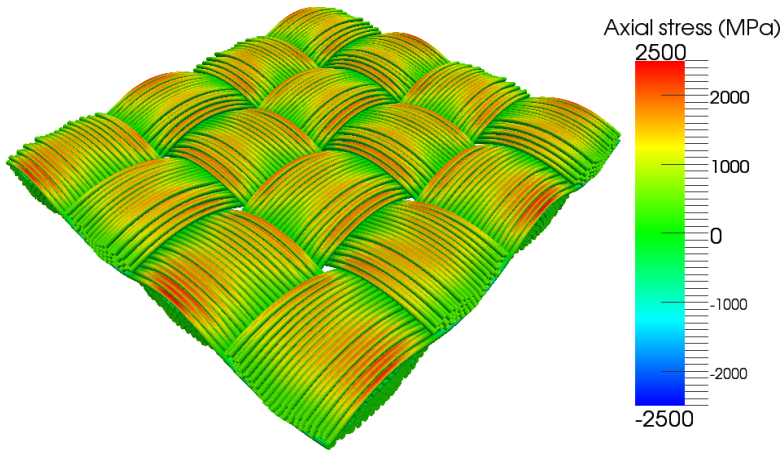

Figure 15.7 : computed initial configuration for the plain weave

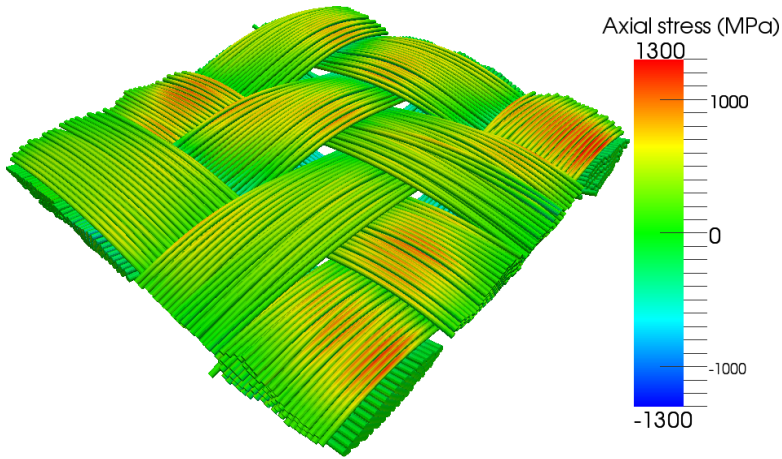

Figure 15.8 : computed initial configuration for the twill weave

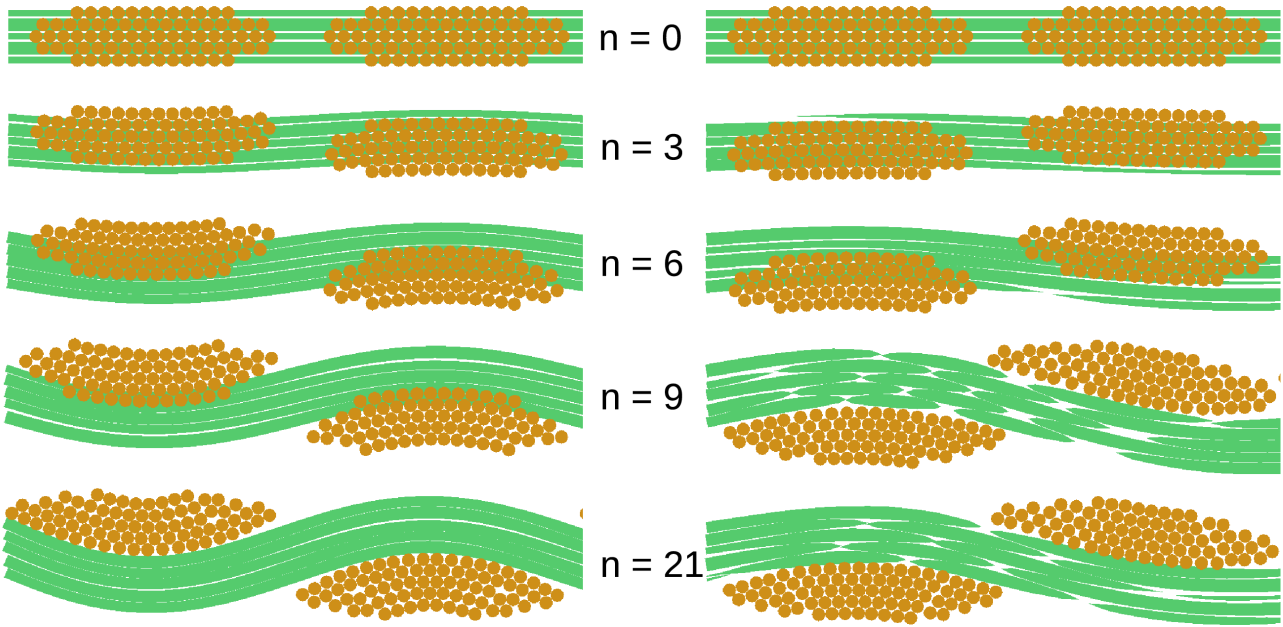

Figure 15.9 : details of the evolution of tows cross-sections for different increments during the computation of the initial configuration for a plain weave (left) and a twill weave (right).

## Plain weave

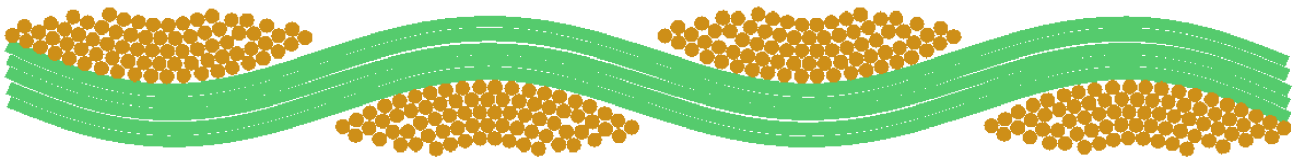

## Twill weave

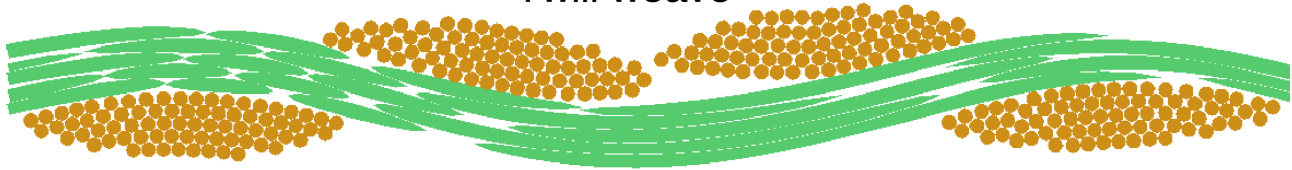

Figure 15.10 : slices of the computed intial configuration for the plain and twill weaves.

### 15.4.2 Application of test loading cases on the dry fabric

Test loading cases can be applied to samples of dry fabric once their initial configuration has been determined.

**Equibiaxial extension**

To simulate an equibiaxial extension on both samples of woven fabric, incremental displacements are applied on the edges of the samples until a 1% strain in warp and weft directions. The loading curves (Fig. 15.11) show a nonlinear effect at the begining of the loading, that may be attributed to changes induced in the arrangement of fibres. Fig. 15.12 and 15.13 show a compaction of fibres within tows, and a reduction of the undulations of tows, mainly in the twill weave.

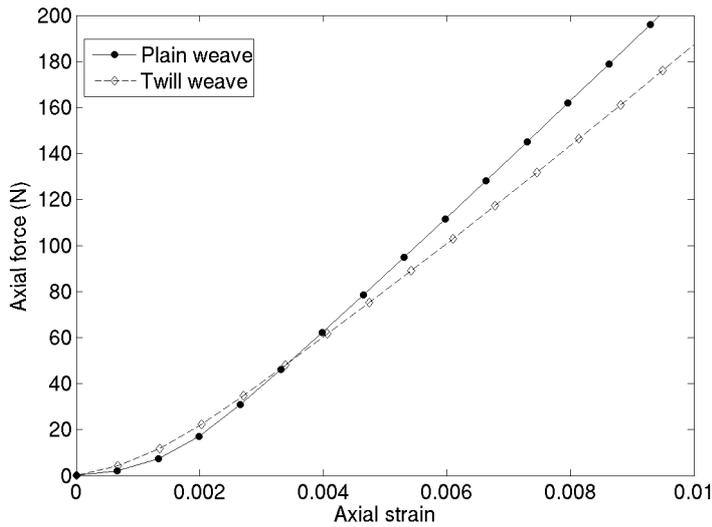

Figure 15.11 : loading curves for the equibiaxial extension

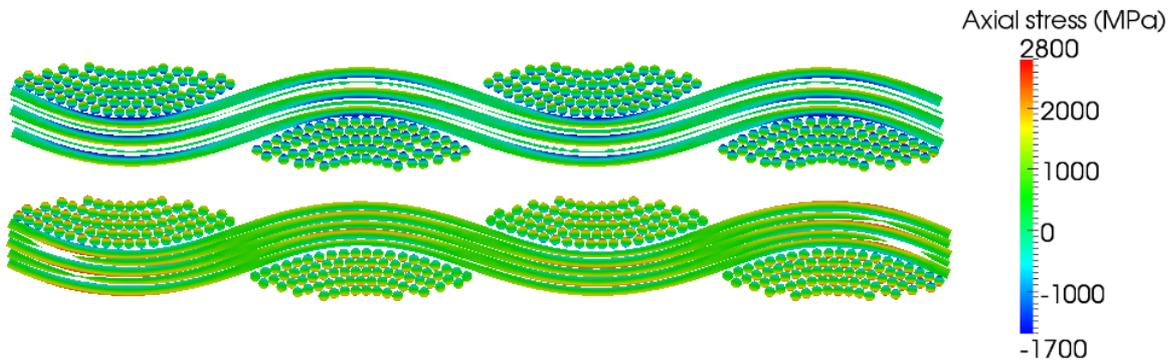

Figure 15.12 : slice of the plain fabric, before and after applying a 1% equibiaxial traction

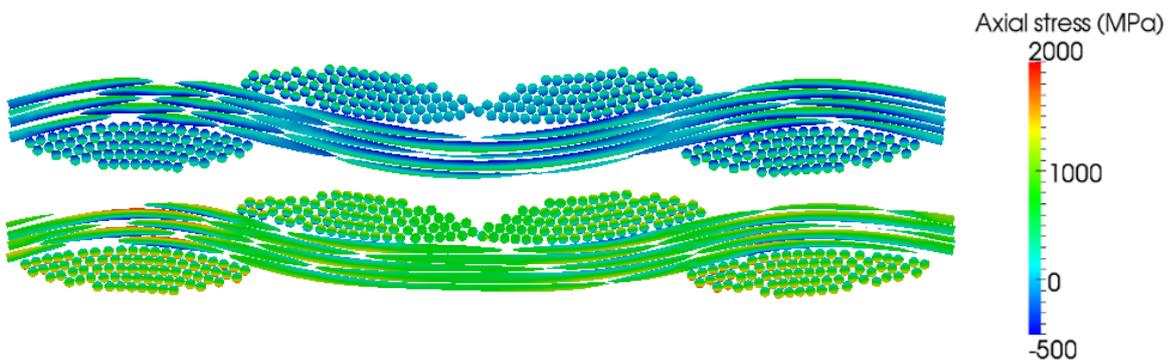

Figure 15.13 : slice of the twill fabric, before and after applying a 1% equibiaxial traction

**Shear test**

A shear test is simulated on the plane weave sample by applying an incremental lateral displacement on one edge of the sample, while exerting a small tensile force in warp and weft directions. The shear loading curve (Fig. 15.14) displays a nonlinear behaviour similar to those observed in benchmark tests on different woven fabrics (Cao *et al.* 2008). The shear force increases until approaching a locking angle for which there is almost no more space between tows , as can be seen in Fig. 15.15 and 15.16.

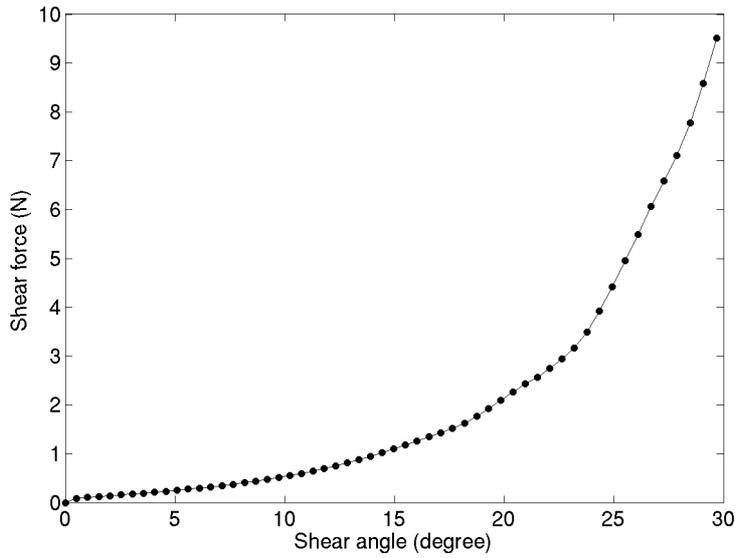

Figure 15.14 : loading curve for the shear test for the plain weave sample

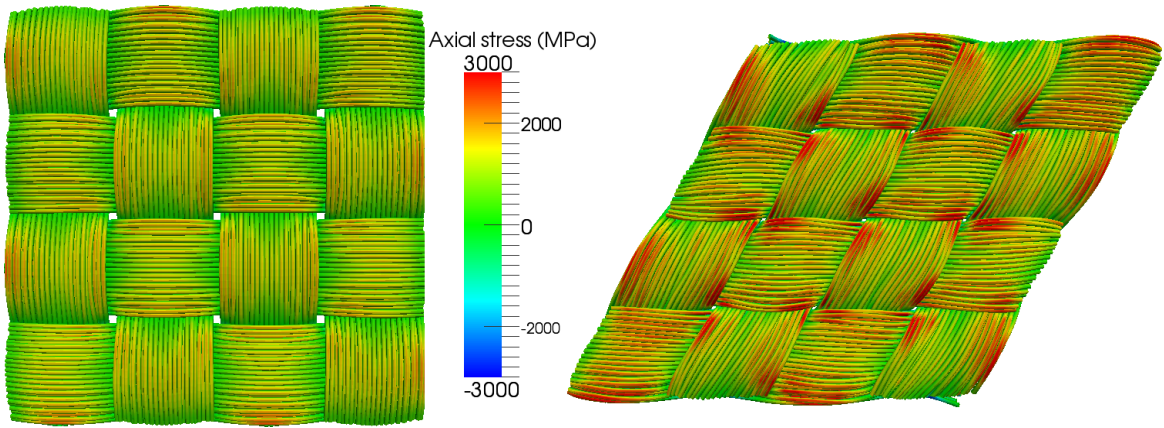

Figure 15.15 : mesh of modeled plain weave sample before and after the application of the shear deformation

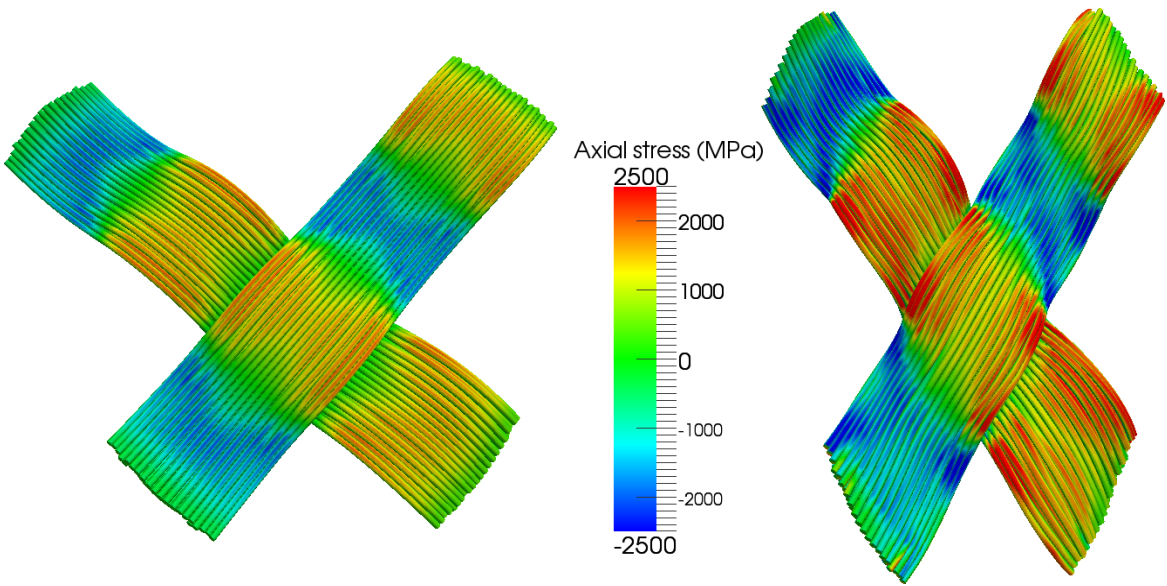

Figure 15.16 : mesh of two particular tows of the plain weave sample before and after the application of the shear deformation

15.4.3 Application of loading cases to the composite sample

Various loading cases can be applied to the composites sample, considering an elastic matrix on both sides of the fabric. The use of rigid bodies to drive boundary conditions enables to apply global conditions on the edge of the sample. Applying for example rotations around different directions while keeping free certain displacements, it is possible to simulate either a global bending (Fig. 15.17 and 15.18) or a twisting (Fig. 15.19). These results demonstrate the ability of the approach to consider large displacements and strains, even in the presence of a matrix. Interesting coupling effects between the matrix and the fabric should be studied by this means.

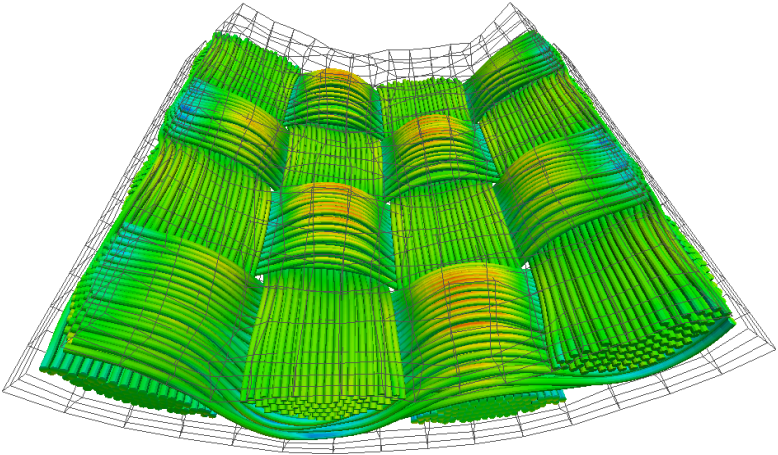

Figure 15.17 : deformed mesh of the plain weave composite sample subject to a global bending

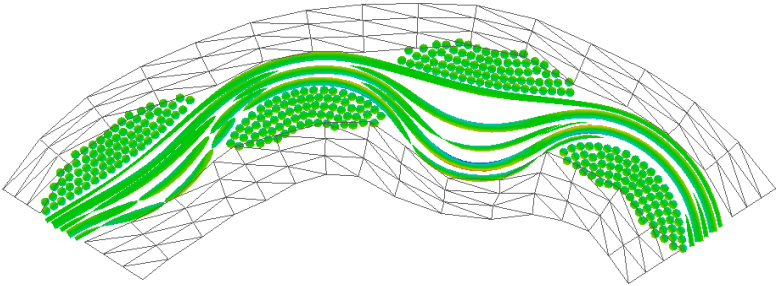

Figure 15.18 : slice of deformed mesh of the plain weave composite sample subject to a global bending

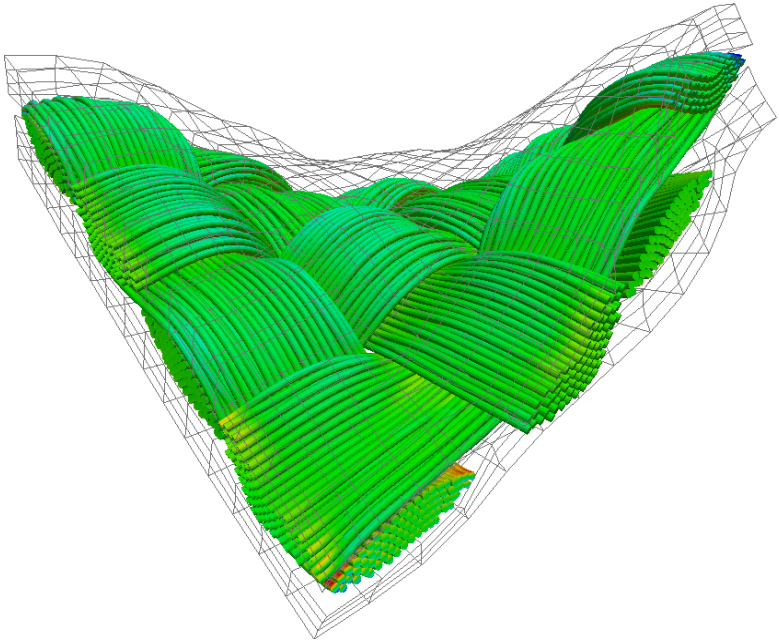

Figure 15.19 : deformed mesh of the plain weave composite sample subject to a global twisting

## 15.5 Conclusions

An approach to the finite element simulation of the mechanical behaviour of textile composite samples at the scale of constituting fibres has been presented in this chapter. Based on tan implicit solver, the approach considers each individual fibre with a finite strain beam model and focuses on the detection and modeling of contact-friction interactions occuring between these fibres. The coupling between fibres and an elastic matrix discretized with a coarse mesh is taken into account by the means of connecting elements generated between the two structures. Appropriate boundary conditions are established on the edges of the studied samples using rigid bodies defined at ends of different components.

The efficiency of the proposed models and algorithms allows to handle samples made of few hundreds of fibres. The approach is first employed to determine the unknown intial configuration of the woven structure, providing geometrical descriptions at the level of fibres and tows in terms of trajectories and cross sections shapes. Once their initial geometry has been computed, various loading cases can then be applied to the studied samples in order to identify their mechanical properties.

This approach at microscopic scale is characterized by the fact it requires the identification of very few parameters (mechanical characteristics of fibres, geometrical description of the weaving pattern), and provides a wide range of results, from the determination of the initial configuration to the characterization of the mechanical state at the scale of fibres during the different loadings. For this reason, it should be helpful to identify behaviours of components at upper scales, both from a geometrical and mechanical point of view.